\documentstyle[epsf,preprint,aps]{revtex}
\tightenlines
\begin{document}
\draft
\title{Thirring solitons in the presence of dispersion}
\author{Alan R. Champneys}
\address{Department of Engineering Mathematics, \\
University of Bristol, Bristol BS8 1TR UK }
\author{Boris A. Malomed}
\address{
Department of Interdisciplinary Studies, \\
Faculty of Engineering, Tel Aviv University, Tel Aviv 69978, Israel}
\author{Mark J. Friedman}
\address{Department of Mathematics, University of Alabama,Huntsville, Alabama}
\date{DRAFT \today }
\maketitle

\begin{abstract}
The effect of dispersion or diffraction on zero-velocity solitons is studied
for the generalized massive Thirring model describing a nonlinear optical
fiber with grating or parallel-coupled planar waveguides with misaligned
axes. The Thirring solitons existing at zero dispersion/diffraction are
shown numerically to be separated by a {\it finite gap} from
three isolated soliton branches. Inside the gap, there is an infinity of
multi-soliton branches. Thus, the
Thirring solitons are {\it structurally unstable}. In another parameter
region (far from the Thirring limit), solitons exist everywhere.
\end{abstract}





The massive Thirring model (MTM) \cite{Thirring} is a completely integrable 
\cite{int} Lorentz-invariant model of classical field theory, which supports
exact soliton and multi-soliton solutions \cite{Thirring}. A generalization
of the MTM, which we call the optical model (OM), was introduced in \cite
{CJ,AW} to describe interaction between right- and left-traveling waves in a
nonlinear optical fiber with a grating. Resonant Bragg scattering and
cross-phase modulation (CPM) give rise, respectively, to linear and
nonlinear couplings between the two waves. OM additionally includes
self-phase-modulation (SPM) nonlinear terms, making it {\it non}integrable
and destroying Lorentz invariance. Nevertheless, a family of {\em exact}
one-solitons can be found \cite{AW,CJ} with arbitrary velocity and internal
amplitude (``soliton'' hereafter means solitary wave, and ``$n$-soliton'' is
one with $n$ peaks). Recently, Bragg solitons have been observed
experimentally in a fiber with grating \cite{exp}.

Both MTM and OM neglect dispersion of the medium, solitons being
supported by an {\em effective dispersion} induced by the linear coupling.
In physical media, however, material dispersion is present. The aim of this
work is to examine the influence of such dispersion $D$ on the Thirring
solitons (TS). This first study treats only zero-velocity solitons,
which are essentially the same in MTM and OM. Results for finite-velocity (
{\em walking} \cite{walking}) solitons, to be presented elsewhere, are more
complicated technically but not drastically different (see below). The
zero-velocity solitons are most intriguing physically, as they imply
complete dynamical self-trapping of light on the grating. We will conclude
that TS are {\it structurally unstable} (which does \/{\em not} imply their
dynamical instability), being separated by a finite gap from the nearest
branch of fundamental-solitons for $D>0$, and with no solitary waves at all
for $D<0$. Within the gap, we find infinite sequences of two-solitons that
are bound states (BS's) of the fundamental ones. Although likely to be
dynamically unstable \cite{BS}, BS's are worth studying to delimit the
existence domain of fundamental solitons, see Figs. 2,4 below.

The generalized MTM including dispersion terms is 
\begin{eqnarray}
iu_t+iu_x+Du_{xx}+\left( \sigma |u|^2+|v|^2\right) u+v &=&0,  \label{PDE1} \\
iv_t-iv_x+Dv_{xx}+\left( |u|^2+\sigma |v|^2\right) v+u &=&0,  \label{PDE2}
\end{eqnarray}
where $u(x,t)$ and $v(x,t)$ are the complex amplitudes of the
counterpropagating waves, $x$ and $t$ are the coordinate and time, $D$ is
the coefficient of spatial dispersion, and $\sigma $ is the relative SPM
coefficient, which is zero for MTM, and $\frac 12$ for the OM case. Besides
fibers with grating, the model (\ref{PDE1}), (\ref{PDE2}) can be applied 
to stationary fields in two parallel tunnel-coupled planar nonlinear
waveguides. In that case $t$ and $x$ are the propagation
distance and the transverse coordinate respectively, 
the terms $\pm iu_x$ account for
misalignment of optical axes in the two cores, $D$ is an effective
diffraction (not dispersion) coefficient, and the CPM terms must be
omitted (see, e.g., \cite{Mak}). Actually, the latter realization of the
model is closer to experiment, as optical axes misalignment is a powerful
control parameter enabling rescaling of physically realistic systems into
the form\ (\ref{PDE1}), (\ref{PDE2}) \cite{Mak}. In contrast, for fibers
with grating, a simple estimate shows that 
dispersion may not be conspicuous unless the spatial width of the soliton is
comparable to the grating period, i.e.\ the wavelength of light, when Eqs.
(1) and (2) are not applicable \cite{wavelength}

Essentially the same model governs interaction of two circular polarizations
of light in a nonlinear fiber, in which the linear coupling is induced by
the birefringence, and the group-velocity difference by a fiber's twist (see
the review \cite{TW}). In untwisted fibers, interaction between linear
polarizations is described by similar models but with linear coupling
replaced by a cubic four-wave-mixing term (see \cite{Spain} for a family of
walking solitons in the latter model).
Our approach is different;
instead of starting from solitons of decoupled nonlinear-Schr\"{o}dinger
equations with the couplings treated as perturbations, we start from the TS
of the strongly coupled system with dispersion or diffraction being a
perturbation.


Being interested here only in the zero-velocity solitons, we substitute into
Eqs.\ (\ref{PDE1}) and (\ref{PDE2}) $u(x,t)=e^{-i\omega t}U(x)$, $
v=e^{-i\omega t}V(x)$ to obtain the coupled ODEs 
\begin{eqnarray}
DU^{\prime \prime }+iU^{\prime }+\omega U+(\sigma |U|^2+|V|^2)U+V &=&0,
\label{ODE1} \\
DV^{\prime \prime }-iV^{\prime }+\omega V+(|U|^2+\sigma |V|^2)V+U &=&0,
\label{ODE2}
\end{eqnarray}
the prime standing for $d/dx$. In this notation, the TS occur at $D=0$ and $
|\omega |<1$. Eqs. (\ref{ODE1}) and (\ref{ODE2}) are equivalent to an
8th-order dynamical system with two integrals of motion: the Hamiltonian 
\begin{eqnarray}
H &=&D(|U^{\prime }|^2+|V^{\prime }|^2)+\omega (|U|^2+|V|^2)+(\sigma
/2)(|U|^4+|V|^4)  \nonumber \\
&&+|U|^2|V|^2+(UV^{*}+VU^{*}),  \label{H}
\end{eqnarray}
and the ``angular momentum'', generated by invariance with respect to the
continuous phase transformation, 
\begin{equation}
M=D\left( UU{^{\prime }}^{*}-U^{*}U^{\prime }+V{V^{\prime }}
^{*}-V^{*}V^{\prime }\right) +|V|^2-|U|^2.  \label{M}
\end{equation}
This Hamiltonian system has several discrete symmetries: the odd symmetry $
Z:(U,V)\to (-U,-V)$, two other 
${\bf Z}_2$ ones $Z_1:U\leftrightarrow V^{*}$, $Z_2:U\leftrightarrow 
-V^{*}$, and four reversibilities 
\begin{eqnarray}
R:(U,U^{\prime },V,V^{\prime }) &\to &(U^{*},-{U^{\prime }}^{*},V^{*},-{
V^{\prime }}^{*}),:x\to -x\,,  \label{R} \\
S:(U,U^{\prime }) &\leftrightarrow &(V,-V^{\prime }),:x\to -x\,,  \label{S}
\end{eqnarray}
along with their odd images $ZR$ and $ZS$.

The first step in locating solitary waves is to solve the linearized
problem, assuming solutions $\sim e^{\lambda x}$. This problem, solved {\em 
exactly}, gives a set of double eigenvalues: 
\begin{equation}
\left( D^2\lambda ^4+2D\omega {\lambda }^2+{\ \lambda }^2+{\ \omega }
^2-1\right) ^2=0\,.  \label{eigen}
\end{equation}
Eq.\ (\ref{eigen}) defines four regions on the plane $\{D,\omega \}$ with
different types of eigenvalues (see Fig.\ 1). Solitary-waves with
exponentially decaying tails are only possible in regions I, II, and III
(and their images for $D<0$), where eigenvalues with nonzero real part occur.

We notice that Eqs.\ (\ref{ODE1}) and (\ref{ODE2}) are compatible with the
reduction $U=V^{*}$. This results in a single equation for $U(x)$, 
\begin{equation}
DU^{\prime \prime }+iU^{\prime }+\omega U+(1+\sigma )|U|^2U+U^{*}=0,
\label{4thODE}
\end{equation}
equivalent to a real fourth-order ODE system. All the zero-velocity solitons
in MTM and OM obey exactly the same reduction, and a simple argument based
on consideration of the unstable manifolds shows that {\em all} possible
zero-velocity solitons to (\ref{PDE1}), (\ref{PDE2}) within region II are
trivially related to solutions of (\ref{4thODE}) by rotation in the $(U,V)$
plane. Henceforth, we set $\sigma =0$ because $\sigma $ can be scaled out
from Eq.\ (\ref{4thODE}). Furthermore, for Eq. (\ref{4thODE}), $S\equiv R$,
and the ``angular momentum'' (\ref{M}) identically vanishes. The eigenvalues
of the corresponding linearized equation are given by Eq.\ (\ref{eigen}),
but are all single, i.e., Fig. 1 remains fully relevant.
\begin{figure}[tbp]
\centerline{\epsfxsize 6.0cm \epsffile{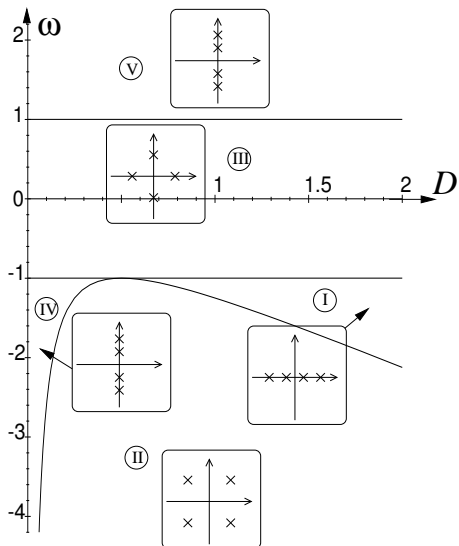}}
\caption{ Parameter regions for $D>0$ with different types of eigenvalues of
the linearized Eqs. (\ref{ODE1}) and (\ref{ODE2}), as illustrated by the
insets. The curve delimiting region II is $D+1/4D-\omega =0$. At the point $
D=\frac 12$, the curve is tangent to the horizontal $\omega =-1$. The
picture for $D<0$ is obtained by rotating the figure by $180^{\circ }$.}
\end{figure}

The soliton is a homoclinic-to-zero solution to Eq. (\ref{4thODE}).
According to general theorems \cite{SadCenter}, in region III, where $U=0$
is a saddle-center fixed point, homoclinic trajectories that are symmetric
under a reversibility are of codimension one (non-symmetric homoclinic
trajectories are of codimension two). Hence solutions can exist only on
isolated curves in the $\{D,\omega \}$ parameter plane, the number of which
may be finite or infinite. Moreover, given a sign condition on the quadratic
part of the Hamiltonian, each curve will be accompanied by an infinite
accumulation of curves on which BS's exist \cite{SadCenter}. In contrast, in
regions I and II, where the fixed point $U=0$ is hyperbolic, homoclinic
trajectories are generic, i.e., they occur uniformly in two-dimensional
parameter regions \cite{SadFocus}. But region III is of most interest, as it
abuts the segment $\{D=0,|\omega |<1\}$ on which the TS solitons exist.

To obtain solutions, we use robust numerical methods for solving two-point
boundary-value problems on a truncation of an infinite $x$-interval with
boundary conditions placing the solution in the stable or unstable
eigenspaces at the origin; see \cite{num} and references therein. 
Continuation of solutions with respect to parameters is
carried out using the software {\sc auto} \cite{auto},
specifically exploit the reversible structure of (\ref{4thODE}). 

Our main findings are summarized in Fig.\ 2. Here, three solid curves
represent the isolated loci of fundamental or {\em primary} (single-humped,
in one component) solitons, and the dashed curves are a small sample of loci
of their two-humped BS's. All primary-solitons are reversible with respect
to the transformation $ZR$, see Eq.\ (\ref{R}); we have found no evidence of
any $R$-reversible solutions. In panel (b), we use, instead of the frequency 
$\omega$, the soliton's energy $E=\int_{-\infty }^{+\infty }|U(x)|^2dx$.
Typical examples of one-solitons are displayed in Fig. 3. 
 and typical two-humped BS's are shown in Fig. 4
(only half of each two-soliton is shown in this figure).

\begin{figure}[tbp]
\centerline{\epsfxsize 3.5in \epsffile{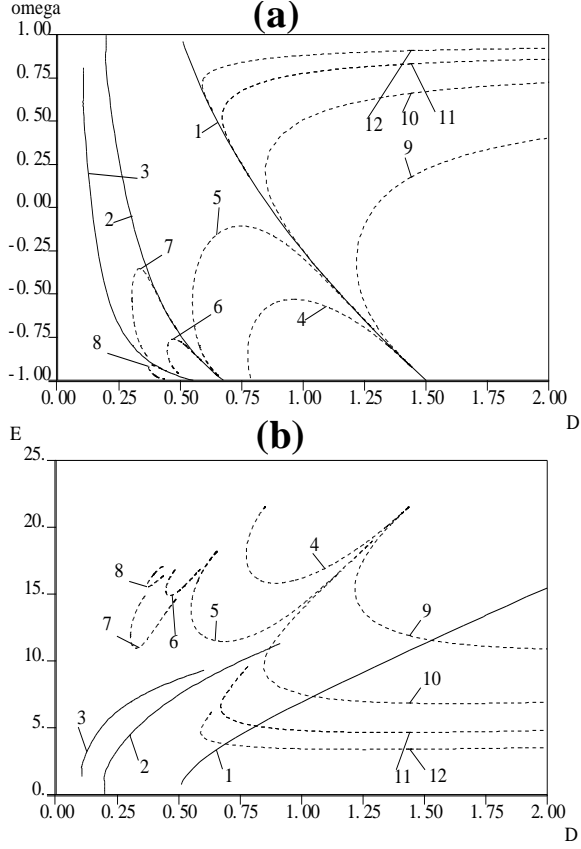}}
\caption{A two-parameter bifurcation diagram for fundamental solitons (solid
curves) and two-soliton bound states (dashed curves) on the planes $
(D,\omega)$ (a) and ($D$,energy) (b).}
\end{figure}
\begin{figure}[tbp]
\centerline{\epsfxsize 3.5in \epsffile{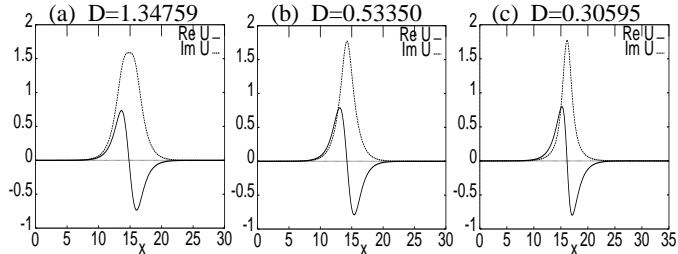}}
\caption{The fundamental solitons at the points of intersection of the
primary-soliton branches (Fig. 2) by the line $\omega =-0.8$.}
\end{figure}

Each of the primary branches in Fig. 2 (labeled 1--3) appears to bifurcate
at zero soliton amplitude from the line $\omega =1$, although there are
numerical difficulties in computing right up to this singular limit. The $D$
-values of these three bifurcations at $\omega =1$ are $D=0.50$, $0.20$ and $
0.11$ to two decimal places. A straightforward calculation of the sign
condition in \cite{SadCenter} on the Hamiltonian (\ref{H}) implies that
curves of $ZR$-reversible BS's must accumulate on each of primary curves
from both sides (e.g. we have found BS branches 9--12 and 4,5 accumulating
on branch 1 from the right and left respectively). Branches 9--12 are also
part of a larger sequence we have computed which for fixed $D$ accumulates
on $\omega =1$.

Three-solitons and higher-order BS of the primary solitons can also be
found, the three-humped ones accumulating on two-solitons, etc. in accord
with the theory \cite{SadCenter}. We do not describe these objects because
it is unlikely that even the two-solitons may be dynamically stable in the
PDE system (\ref{PDE1}),(\ref{PDE2}), while stability of the primary
solitons is quite feasible \cite{BS}. However, stability analysis is
deferred to another work. Homoclinic solutions were also sought for $D<0$
and $|\omega |<1$, but no evidence of primary or multi-humped ones was found.

Looking at Fig.\ 2, there remains the crucial question whether there
are any more primary branches to the left of that labeled 3.  A
seemingly plausible conjecture is that there is a self-similar
structure of primary branches as one moves to the left in Fig. 2,
i.e., infinitely many branches accumulating on the TS segment
$\{D=0,|\omega |<1\}$, the branches 1,2 and 3 being but the first
three in the structure (note that, at least for $D<2$, there cannot be
any further primary solutions to the {\it right} of branch 1, because
here the two-soliton curves 9--12 form a barrier for them). However,
careful numerical scanning of the parametric plane of Fig. 2 to the
left of branch 3 has strongly indicated that the above hypothesis is
{\it false}, in region III there being {\it no} primary branches other
than $1$, $2$, and $3$. For $\omega $ sufficiently close to $-1$, this
assertion is substantiated as follows

Fig 4.\ reports the results of a thorough numerical investigation of other
possible solution branches at $\omega =-0.99$, varying $D$ between $0$ and $
\frac 12$. We find that, to the left of branch 3, an infinite sequence of 
{\it multi}-soliton BS's occurs. Even though, because of numerical problems
in the singular limit, we have only computed the corresponding two-solitons
down to $D\approx 0.2$, Fig. 4 clearly suggests accumulation of the sequence
as $D\to 0$. The energy of the two-solitons remains finite, while the
separation between the two bound pulses diverges $\sim 1/D$ as $D\to 0$
(which explains the existence of TS in the limit $D=0$). Thus, what does
accumulate on the TS manifold at $D\to 0$ is an infinite sequence of
multi-soliton branches, 
with no fundamental-soliton branch closer to the TS manifold than the branch
3 in Fig. 2.

To support this numerical finding with qualitative arguments, consider what
happens to the primary branches as they cross the line $\omega =-1$ from
above. For $D>\frac 12$ this is a ``harmless'' transition, because the real
eigenvalues of the linearized equations, that govern the decay of the
homoclinic solution at $|x|\to \infty $, behave smoothly and they are
bounded away from zero. 
A well-defined primary branch safely crosses $\omega =-1$ in this case,
which for $\omega <-1$ describes a curve of ``orbit-flip'' bifurcations
(cf.\ \cite{OrbitFlip}). However, for $D<\frac 12$, the corresponding
eigenvalues vanish as $\omega \to -1$, hence no smooth transition can take
place. Thus, there may be {\em no} primary-soliton branches at $0<1+\omega
\ll 1,\;D<\frac 12$.

\begin{figure}[tbp]
\centerline{\epsfxsize 3.5in \epsffile{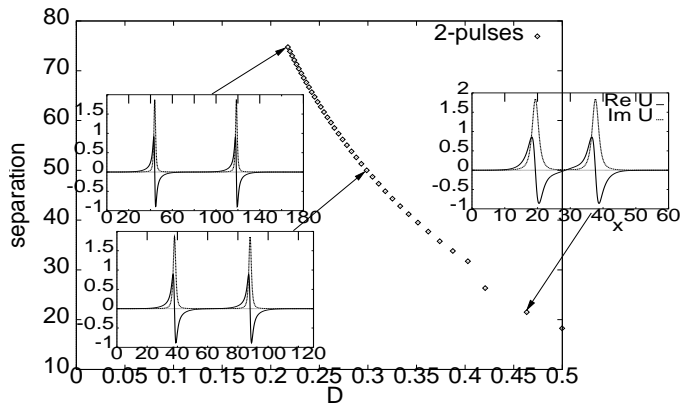}}
\caption{A sequence of two-soliton solutions at $D\to 0$ for $\omega=-0.99$.
The insets show the shape of the solutions.}
\end{figure}
The most important result of this work is that there is a {\it finite gap}
separating TS, existing in the singular limit $D=0$, from new solitons at $
D\neq 0$. Thus, the Thirring solitons are {\it structurally unstable}
against adding the dispersion or diffraction. A natural question is if there
is a gap for solitons at a finite velocity. Preliminary numerical results
give a positive answer, which is further supported by an argument that
solutions to the ODEs describing the soliton's shape continuously depend on
the parameters (including velocity), except at a singular point. The
addition of dispersion to MTM is, obviously, a singular perturbation because
it doubles the system's order; however, nonzero velocity is not a singular
perturbation.

We mention finally results for solitons in regions I and II. As stated, here
homoclinic solutions are generic, and a primary soliton branch can be 
path-followed continuously for {\it all} $\omega $ and $D$ inside regions I
and II. Inside region II it develops oscillations in its tails due to the
complex eigenvalues. 
At the boundaries between regions I and III and II and IV, the solution
disappears through a zero-amplitude bifurcation, as predicted by the
appropriate normal-form analysis \cite{Io}. Other primary-soliton solutions
have more complicated bifurcation diagrams; in both regions I and 
II, two- and multi-soliton BS's also occur. A detailed
description of the complete bifurcation structure will be given elsewhere.

Since the original submission of this paper, we have become aware of
the preprint \cite{BPZ}, containing new results on the {\em dynamical}
stability of the solitons in OM {\em without} the dispersion terms.
They demonstrate that, except for the integrable Thirring model case,
all the solitons are subject to an instability which is too weak to
have been observed in earlier numerical simulations.  Note that a
similar instability mechanism for solitons of OM was predicted
non-rigourously in \cite{MT} using a variational approximation.  A
dynamical stability analysis for the new solitons in the presence of
dispersion found in the present work will be presented elsewhere.

We appreciate valuable discussions with Y.S. Kivshar, G.G. Luther
and D.E. Pelinovsky.

\end{document}